\documentclass[aps,pra,eqsecnum,twocolumn,showpacs]{revtex4}

\begin{document}
\title{Optimum unambiguous discrimination of two mixed states and
application to a class of similar states}
\author{Ulrike Herzog}
\affiliation {Institut f\"ur Physik, Humboldt-Universit\"at Berlin, Newtonstrasse 15, D-12489 Berlin, Germany}
\date{\today}
\begin{abstract}
We study the measurement for the unambiguous discrimination of two mixed quantum states that are described by
density operators $\rho_1$ and $\rho_2$ of rank $d$, the supports of which jointly span a $2d$-dimensional
Hilbert space. Based on two conditions for the optimum measurement operators, and on a canonical representation
for the density operators of the states, two equations are derived that allow the explicit construction of the
optimum measurement, provided that the expression for the fidelity of the states has a specific simple form. For
this case the problem is mathematically equivalent to distinguishing pairs of pure states, even when the density
operators are not diagonal in the canonical representation. The equations are applied to the optimum unambiguous
discrimination of two mixed states that are similar states, given by $\rho_2= U\rho_1 U^{\dag}$, and that belong
to the class where the unitary operator $U$ can be decomposed into multiple rotations in the $d$ mutually
orthogonal two-dimensional subspaces determined by the canonical representation.

\end{abstract}
\pacs{PACS:03.67.Hk,03.65.Ta,42.50.-p} \maketitle

\section{Introduction}

The discrimination of nonorthogonal quantum states is of fundamental interest for many problems connected with
quantum communication and quantum information and has consequently attracted a great deal of attention. An
overwiew of the theoretical aspects of quantum state discrimination is given in recent review articles
\cite{chefles,springer}. The standard problem is the following: We assume that a quantum system is prepared in a
certain state that is drawn with known prior probability from a finite set of known possible states, and we want
to find the best measurement for determining the actual state of the system. When the given states are
nonorthogonal, they cannot be discriminated perfectly, and therefore various measurement strategies have been
developed that are optimized with respect to different criteria. The most prominent of these schemes are
discrimination with minimum error \cite{helstrom} on the one hand, and optimum unambiguous discrimination
\cite{ivan,jaeger} on the other hand, and very recently also the strategy of discrimination with maximum
confidence has been introduced \cite{croke}.

In a measurement for unambiguous discrimination errors are not allowed, at the expense of admitting inconclusive
results, where the measurement fails to give a definite answer.  In this paper we restrict ourselves to
considering only two given states that in the most general case are mixed states. Clearly, error-free
discrimination of mixed states is only possible when the supports of the states are not identical. Note that the
support of a quantum state is the Hilbert space spanned by those eigenvectors of its density operator that belong
to nonzero eigenvalues, and the rank of a state is the dimension of its support. The optimum error-free
measurement we are trying to find is the measurement that minimizes the average probability of getting an
inconclusive result, or in other words, the average failure probability, where the prior probabilities for the
occurrence of the different possible states are taken into account. We mention that recently unambiguous
discrimination was also investigated without considering these prior probabilities, by requiring that in the
best measurement the largest state-selective failure probability for any of the incoming states be as small as
possible \cite{minimax}. Here we stick to the traditional way of defining optimality for unambiguous
discrimination by requiring that the average overall failure probability of the discriminating measurement be as
small possible.

While the optimum measurement for the unambiguous discrimination of two pure states was found already a long
time ago \cite{ivan,jaeger}, unambiguous discrimination involving mixed states, or sets of pure states,
equivalently, became an object of research only more recently [8--20]. So far a general analytical solution for
the optimum measurement that unambiguously distinguishes between two arbitrary mixed states  does not exist yet,
but a number of general results have been obtained. Several necessary and sufficient conditions for the optimum
measurement have been derived \cite{zhang,eldar}, and it has been shown that the solution can be found in an
efficient way using the method of semi-definite programming \cite{eldar}. Moreover, reduction theorems have been
developed \cite{raynal,raynal2} that can simplify the discrimination problems.

Lower bounds for the failure probability \cite{rudolph,feng1,zhou} as well as the conditions for saturating the
bounds \cite{HB,raynal1,zhou} have been also studied. It has been established that the minimum failure
probability in the unambiguous strategy is at least twice as large as the minimum probability to get a wrong
result when errors are allowed to occur \cite{HB1}. As a consequence of the reduction theorems [8] it follows
that the overall lower bound of the failure probability, proportional to the fidelity \cite{rudolph}, can only
be saturated when the dimension of the joint Hilbert space spanned by the supports of the two states is equal to
the sum of their ranks. Even for this case the saturation of the bound depends sensitively on the structure of
the density operators and on their prior probabilities, and it is expected \cite{BFH,zhou} that in general the
fidelity bound can be reached only in a very limited range of all parameters.

In a few special cases a complete analytical solution for the optimum measurement, valid for arbitrary
prior probabilities of the two given states, has been derived. These cases are the unambiguous discrimination of\\
 { i})  a pure
state and an arbitrary mixed state, known as quantum state filtering \cite{SBH,BHH}, \\
{ ii})  two density operators of rank $d$ in a $(d+1)$-dimensional joint Hilbert space
 \cite{rudolph}, and \\
{ iii})  two density operators of rank $d$ that are simultaneously diagonal in the canonical basis
\cite{rudolph} that separates the 2$d$-dimensional joint Hilbert space into $d$ mutually orthogonal
two-dimensional subspaces
\cite{BFH}.\\
 The solution derived in Ref. \cite{BFH} includes the unambiguous discrimination of two uniformly mixed states,
 which is equivalent
 to the discrimination of the two subspaces spanned by their supports.
  Moreover, the case iii) also applies to the comparison of two given pure states \cite{jex} having
arbitrary prior probabilities \cite{HB,kleinmann} and to the programmable discrimination \cite{BH} that
distinguishes between two pure states when one \cite{BBFHH} or both of them \cite{BH,BBFHH,hayashi} are unknown
 and the discrimination is performed with the help of reference copies.

In the three cases listed above the optimum measurement can be constructed from the solutions for discriminating
pairs of pure states. When such a reduction to pure-state discrimination problems is not possible, two classes
of analytical solutions are known. First, general expressions for the optimum measurement operators have been
derived which hold in the special case that for the given prior probabilities of the states the lower bound of
the failure probability is saturated \cite{raynal1}. Second, the optimum measurement has been determined for two
equally probable geometrically uniform states of rank two, described by density operators $\rho_1$ and $\rho_2$
with $\rho_2= U\rho_1 U$, where $U^2=I$ with $I$ being the identity \cite{raynal2}.

In the present paper we extend the above list of cases where a complete solution, valid for arbitrary prior
probabilities of the states, can be obtained. We show that when a certain condition for the fidelity of the two
states is fulfilled, the solution of the case { iii}) can be used to solve the optimization problem also for
states that are not diagonal when represented with the help of the respective sets of orthonormal canonical
basis states. In particular, the required condition for the fidelity is found to be satisfied for two mixed
states that are similar, given by $\rho_2= U\rho_1 U^{\dag}$ where $U^{\dag}U = I$, and that in addition belong
to the class where the unitary operator $U$ can be decomposed into multiple rotations in the two-dimensional
subspaces determined by the canonical representation of the mixed states. Interestingly, by extending the
quantum key distribution protocol based on two nonorthogonal pure states \cite{bennett} to the case of two mixed
states, it has been found a decade ago \cite{koashi} that secure communication is only possible in this protocol
when the two mixed states are connected by a rotation operator with a nonorthogonal angle and belong to the
special class of states considered in this paper.

The paper is organized as follows: In Sec. II we review earlier results that are needed for the present
investigation, and we derive our basic equations. The optimum measurement for the unambiguous discrimination of
the special class of states considered in this paper is obtained in Sec. III, and the paper is concluded in Sec.
IV.

\section{General theory}

\subsection{Conditions for the lower bound of the failure probability}

We start with a brief summary of the basic theoretical concepts and results that are needed for our further
treatment. Any measurement for distinguishing two quantum states, characterized by the density operators
$\rho_1$ and $\rho_2$,
  can be formally described by three positive detection operators obeying the equation
\begin{equation}
\label{pi} \Pi_0 + \Pi_1 + \Pi_2 = I,
\end{equation}
where $I$ is the identity. These detection operators are defined in such a way that $\rm Tr(\rho\Pi_k)$
 with $k=1,2$ is the probability that a system
prepared in a state $\rho$ is inferred to be in the state $\rho_k$, while $\rm Tr(\rho\Pi_0)$ is the probability
that the measurement fails to give a definite answer. The measurement is a von Neumann measurement when all
detection operators are composed of projectors, otherwise it is a generalized measurement based on a positive
operator-valued measure (POVM). From the detection operators $\Pi_k$ schemes for realizing the measurement can
be obtained using standard methods \cite{preskill}. For the results of the measurement to be unambiguous, errors
are not allowed to occur so that there is never a misidentification of any of the states. This leads to the
requirement
\begin{equation}
\label{unamb} \rho_1 \Pi_2 = \rho_2 \Pi_1= 0
\end{equation}
\cite{chefles,springer}, which means that
$\rm Tr (\rho_k\Pi_0)= 1 - \rm Tr (\rho_k\Pi_k)$ for $k=1,2$.
When we denote the prior probabilities for the occurrence of the two states by $\eta_1$ and $\eta_2$,
respectively, with $\eta_1 + \eta_2 = 1$, the total failure probability of the measurement, Q, is given by
\begin{eqnarray}
\label{Q} Q &=& \eta_1 \rm Tr (\rho_1\Pi_0) + \eta_2 \rm Tr (\rho_2\Pi_0)\nonumber\\
&=&1-\eta_1 \rm Tr (\rho_1\Pi_1) - \eta_2 \rm Tr (\rho_2\Pi_2).
\end{eqnarray}
 From the relation between the arithmetic and the geometric mean
 we get
$Q     \geq   2\sqrt{\eta_1 \eta_2
           \rm Tr(\rho_1\Pi_0)\rm Tr(\rho_2\Pi_0)}$,
           and because of the Cauchy-Schwarz-inequality this yields
 $Q \geq  2\sqrt{\eta_1 \eta_2}\,
 {\rm Max}_V\, |{\rm Tr}(V\sqrt{\rho_1}\Pi_0\sqrt{\rho_2})|$
\cite{feng1}, where $V$ describes an arbitrary unitary transformation. The failure probability takes its
absolute minimum when the equality signs hold in these two relations, which is true if and only if both the
equations
\begin{equation}
\label{tr} \eta_1 \rm Tr(\rho_1\Pi_0)= \eta_2 \rm Tr(\rho_2\Pi_0)
\end{equation}
and $V \sqrt{\rho_1}\sqrt{\Pi_0} \sim \sqrt{\rho_2}\sqrt{\Pi_0}$
 are fulfilled.
After multiplying the second relation with its Hermitean conjugate, the two conditions for equality can be
combined to yield \cite{HB}
\begin{equation}
\label{cond} \eta_1\sqrt{\Pi_0} \rho_1 \sqrt{\Pi_0}  =  \eta_2\sqrt{\Pi_0} \rho_2 \sqrt{\Pi_0}.
\end{equation}
Substituting  $\Pi_0=I - \Pi_1 - \Pi_2$ into the inequality for the failure probability $Q$, given above, we
arrive at
\begin{equation}
\label{Q0}
Q   \geq
    2\sqrt{\eta_1 \eta_2}\; F(\rho_1,\rho_2),
\end{equation}
where
\begin{equation}
\label{F} F={\rm Tr}\,[\left(\sqrt{{\rho}_2}\; {\rho}_1 \sqrt{{\rho}_2}\right)^{1/2}]={\rm
Tr}|\sqrt{\rho_1}\sqrt{\rho_2}|
\end{equation}
 is the fidelity \cite{nielsen}.
From Eqs. (\ref{Q}) and (\ref{tr}) we conclude  that the lower bound of the failure probability, proportional to
the fidelity of the states, is obtained if and only if $\eta_1 {\rm Tr}(\rho_1\Pi_0) = \eta_2 {\rm
Tr}(\rho_2\Pi_0)=\sqrt{\eta_1 \eta_2}\,F$. This is equivalent to the two conditions \cite{HB}
\begin{eqnarray}
\label{cond1}
 {\rm Tr}(\rho_1\Pi_1) -  1+ \sqrt{\frac{\eta_2}{ \eta_1}}\,F(\rho_1,\rho_2)&=& 0, \\
\label{cond2}
 {\rm Tr}(\rho_2\Pi_2) -1 +  \sqrt{\frac{\eta_1}{ \eta_2}}\,F(\rho_1,\rho_2)&=& 0
\end{eqnarray}
that are the basic equations for our further treatment. Whenever we can find detection operators $\Pi_1$ and
$\Pi_2$ satisfying Eqs. (\ref{cond1}) and  (\ref{cond2}) while $\Pi_0 = I -\Pi_1 -\Pi_2$ is also a detection
operator, i. e. a positive operator with eigenvalues between 0 and 1, then we are sure that these operators
determine the optimum measurement for unambiguously  discriminating the states, since they yield the lower bound
of the failure probability, proportional to the fidelity. In the optimum measurement the lower bound can only be
achieved when the necessary, but not sufficient, condition \cite{HB}
\begin{equation}
\frac{{\rm Tr}(P_2 \rho_1)}{F} \leq \sqrt{\frac{\eta_2}{\eta_1}}
        \leq \frac{F}{{\rm Tr}(P_1 \rho_2)}
\label{ness-cond}
\end{equation}
is fulfilled, where the operators $P_1$ and $P_2$ are the projectors onto the supports of $\rho_1$ and $\rho_2$,
respectively. It has  been pointed out that there exist mixed states for which the fidelity bound cannot be
reached for any value of the prior probabilities \cite{HB,BFH}.

\subsection{The canonical representation of the density operators}

When we want to explicitly determine the optimum detection operators, it is crucial to use convenient basis
vectors for representing the two given states. From now on we focus our interest to the  problem of
distinguishing two states of rank $d$ the supports of which jointly span a $2d$-dimensional Hilbert space,
because it has been shown that the unambiguous discrimination of two arbitrary states  can be reduced to this
standard problem \cite{raynal}.
 We start from the spectral representations for the two given states,
\begin{equation}
\label{rho} \rho_1 = \sum_{i=1}^{d} \tilde{r}_{i} |\tilde{r}_i\rangle\langle \tilde{r}_i|, \qquad \rho_2 =
\sum_{i=1}^{d} \tilde{s}_{i}|\tilde{s}_i\rangle\langle \tilde{s}_i|.
\end{equation}
The projectors onto the supports of the states then read
\begin{equation}
\label{support} P_1 = \sum_{i=1}^{d} |\tilde{r}_i\rangle\langle \tilde{r}_i|, \qquad P_2 = \sum_{i=1}^{d}
|\tilde{s}_i\rangle\langle \tilde{s}_i|.
\end{equation}
As will become obvious later, for our purposes it is advantageous to perform two separate unitary basis
transformations in the two Hilbert spaces ${\cal H}_1$ and ${\cal H}_2$ spanned by the supports of $\rho_1$ and
$\rho_2$, respectively, yielding two new sets of orthonormal basis states that are denoted by $\{|r_i\rangle\}$
and $\{|s_i\rangle\}$ and have the property that
\begin{eqnarray}
\label{basis1a}
\langle r_i|r_{j}\rangle  &=& \langle s_i|s_j\rangle= \delta_{ij},\\
\label{basis1b}
 \langle r_i|s_j\rangle & =&
\langle s_j|r_i\rangle=C_i\delta_{ij}, \quad 0\leq C_i\leq 1.
\end{eqnarray}
Basis states of this kind have been used already previously to study the unambiguous discrimination of two mixed
states \cite{rudolph,BFH} and to construct a very simple example \cite{HB}. After the basis transformations have
been performed, the density operators take the form
\begin{equation}
\label{rho1} \rho_1 = \sum_{i,j=1}^{d} r_{ij} |r_i\rangle\langle r_j|, \qquad \rho_2 = \sum_{i,j=1}^{d}
s_{ij}|s_i\rangle\langle s_j|.
\end{equation}
In the following we shall refer to Eqs. (\ref{rho1}) together with Eqs. (\ref{basis1a}) and (\ref{basis1b})  as
the canonical representation of the two given density operators.

In order to show that for any two density operators of rank $d$ jointly spanning a $2d$-dimensional Hilbert
space the canonical representation always exists, and to give also a recipe how it can be constructed, we rely
on the treatment given in Ref. \cite{koashi}. First we observe that the operator $P_1P_2P_1$ is Hermitean, and
that its eigenstates, which we denote by $|r_i\rangle$, therefore span a complete $d$-dimensional orthonormal
basis in ${\cal H}_{1}$. Because $P_2$ and $P_1$ are projectors, it follows that $\langle r_i| P_2^2 |r_i
\rangle = \langle r_i| P_1P_2P_1 |r_i\rangle$. Clearly, the norm of the state $P_2|r_i\rangle$ is
 not larger than 1, and moreover it is non-zero since the joint Hilbert space spanned by the supports of the two
 density operators is assumed to be $2d$-dimensional. Hence we can establish the eigenvalue equation
\begin{equation}
\label{basis1} P_1P_2P_1 |r_i\rangle=P_1P_2 |r_i\rangle= C_i^2 |r_i\rangle,
 \end{equation}
where $0 < C_i \leq 1$ and $\langle r_i|r_j\rangle = \delta_{ij}$. Now we introduce the normalized states in
${\cal H}_{2}$ that are given by \cite{koashi}
\begin{equation}
\label{basis2} |s_j\rangle = \frac{1}{C_j}\;P_2 |r_j\rangle= \frac{1}{C_j}\;P_2 P_1|r_j\rangle
 \end{equation}
 and obey the equations
\begin{eqnarray}
\label{basis4a} \langle s_i|s_j\rangle & = & \frac{1}{C_i C_j}\;\langle r_i |P_1P_2P_1| r_j\rangle, \\
\label{basis5} \langle r_i|s_j\rangle & = & \frac{1}{C_j}\;\langle r_i |P_1P_2 P_1| r_j\rangle.
 \end{eqnarray}
Taking into account that $\langle r_i |P_1P_2 P_1| r_j\rangle = C_i^2 \delta_{ij}$ because of Eq.
(\ref{basis1}), we immediately arrive at Eqs. (\ref{basis1a}) and (\ref{basis1b}). Thus we have shown that  Eq.
(\ref{basis1}) together with Eq. (\ref{basis2})
 provides the means for determining the two  sets of canonical basis states
$\{|r_i\rangle\}$ and $\{|s_i\rangle\}$. Obviously, this requires the solution of a $d$th-order algebraic
equation, resulting  from the eigenvalue equation,  Eq. (\ref{basis1}).
 We still remark that by making use of   $P_2|r_i\rangle= C_i |s_i\rangle$ and $\quad P_1|s_i\rangle= C_i
|r_i\rangle$, Eq. (\ref{basis1}) can be transformed into $C^2_i\;P_2|r_i\rangle=P_2P_1P_2|r_i\rangle$ which,
with the help of Eq. (\ref{basis2}), leads to the alternative eigenvalue equation $ P_2P_1 |s_i\rangle=
P_1P_2P_1 |s_i\rangle= C_i^2|s_i\rangle$, as expected for symmetry reasons.

\subsection{Construction of the optimum detection operators}

Having obtained the canonical representation of the density operators to be discriminated, we are now in the
position to make an explicit general Ansatz for the detection operators $\Pi_1$ and $\Pi_2$ that enable the
unambiguous discrimination by satisfying Eq. (\ref{unamb}). For this purpose we define the states
\begin{equation}
\label{v} |v_i\rangle = \frac{|r_i\rangle - C_i   |s_i\rangle}{S_i}, \quad
 |w_i\rangle = \frac{|s_i\rangle - C_i  |r_i\rangle}{S_i},
\end{equation}
where
$\label{sin} S_i= \sqrt{1-C_i^2}$ .
Making use of  Eqs. (\ref{basis1a}) and (\ref{basis1b}) it follows that
\begin{eqnarray}
\label{basis3} \langle v_i|v_{j}\rangle  =  \langle w_i|w_j\rangle= \delta_{ij}
\end{eqnarray}
and, most importantly,
\begin{eqnarray}
\label{basis3a}
 \langle v_i|s_j\rangle  = \langle w_i|r_j\rangle = 0.
\end{eqnarray}
The two joint sets of states
 $\{\{|s_i\rangle\},\{|v_i\rangle\}$, on the one hand, and $ \{\{|r_i\rangle\},
\{|w_i\rangle\}\}$, on the other hand, form two different complete orthonormal basis systems in our
2$d$-dimensional Hilbert space. Their mutual geometrical orientation
 is characterized by the relations
\begin{eqnarray}
\label{basis4} \langle v_i|r_j\rangle  =   \langle w_i|s_j\rangle  = S_i\delta_{ij},
\end{eqnarray}
in addition to  $\langle v_i|w_j\rangle  = - C_i\,\delta_{ij}= - \langle r_i|s_j\rangle  $.
 In accordance with our earlier
work \cite{HB} we can now make the general Ansatz
\begin{equation}
\label{detect} \Pi_1 =\sum_{i,j=1}^{d} \alpha_{ij}|v_i\rangle\langle v_j|, \quad
 \Pi_2 =\sum_{i,j=1}^{d} \beta_{ij}|w_i\rangle\langle w_j|
\end{equation}
which because of Eqs. (\ref{rho1}) and (\ref{basis3a}) guarantees that $\rho_1\Pi_2=\rho_2\Pi_1=0$, as required
for unambiguous discrimination. For these operators to describe a physical measurement, the coefficients
$\alpha_{ij}$ and $\beta_{ij}$ must be chosen in such a way that their eigenvalues, as well as the eigenvalues
of $\Pi_0$, are nonnegative and not larger than 1. Using the expression $\Pi_1 =\sum_{i,j}
\alpha_{ij}I|v_i\rangle\langle v_j|I$, where
\begin{equation}
\label{unit} I = \sum_{i=1}^d \left(|r_i\rangle \langle r_i| + |w_i\rangle \langle w_i|\right)
\end{equation}
is the unity operator in the 2$d$-dimensional Hilbert space, we can represent the operator $\Pi_0=I-\Pi_1-\Pi_2$
in the form
\begin{eqnarray}
\label{pi0} \Pi_0 = \sum_{i,j=1}^d [(\delta_{ij} - \alpha_{ij}S_iS_j)|r_i\rangle\langle r_j|
                   +\alpha_{ij}S_i C_j|r_i\rangle\langle w_j|\nonumber\\
                \;\;   +\alpha_{ji}S_j C_i|w_i\rangle\langle r_j|
                   +(\delta_{ij} - \alpha_{ij}C_iC_j-\beta_{ij})|w_i\rangle\langle w_j|].\nonumber\\
\end{eqnarray}
Moreover, from  Eqs. (\ref{detect}) and (\ref{Q}) we obtain an explicit expression for the failure probability,
given by
\begin{equation}
\label{Q2} Q= 1- \sum_{i,j=1}^d S_i S_j (\eta_1 \alpha_{ij} r_{ji}+\eta_2 \beta_{ij}s_{ji}).
\end{equation}
For brevity, in the rest of the paper we denote the diagonal elements of the density operators and of the
detection operators as
\begin{equation}
\label{abbr} r_{ii}\equiv r_i, \quad  s_{ii}\equiv s_i, \quad \alpha_{ii}\equiv \alpha_i, \quad \beta_{ii}\equiv
\beta_i.
\end{equation}
Since $\sum_i r_{i} = \sum_i s_{i} = 1$, the conditions for the achievement of the absolute minimum of the
failure probability, Eqs. (\ref{cond1}) and (\ref{cond2}), can be rewritten as
\begin{eqnarray}
\label{cond3}
  \sum_{i,j=1}^d (S_i S_j \alpha_{ij}-\delta_{ij}) r_{ji}
  +  \sqrt{\frac{\eta_2}{ \eta_1}} F(\{C_i,r_{ij},s_{ij}\}) =0,\;\;\;\;\;\\
\label{cond4}
 \sum_{i,j=1}^d (S_i S_j  \beta_{ij}-\delta_{ij})s_{ji}
 + \sqrt{\frac{\eta_1}{ \eta_2}} F(\{C_i,r_{ij},s_{ij}\}) =0,\;\;\;\;\;
\end{eqnarray}
where the fidelity depends on the parameters that characterize the density operators in the canonical
representation, given by Eqs. (\ref{basis1a})-(\ref{rho1}). Clearly, in general the coefficients $\alpha_{ij}$
and $\beta_{ij}$ are not uniquely determined by these two equations alone, and a complete system of equations
would have to be found, taking into account Eq. (\ref{cond}). However, under certain conditions  Eqs.
(\ref{cond3}) and (\ref{cond4}) are sufficient for obtaining the optimum measurement, as we shall see in the
following. In particular, this is the case when the canonical representation of the density operators is such
that the expression for the fidelity has a specific form, depending only  on the diagonal elements $r_i$ and
$s_i$.

\subsection{An analytical solution for the optimum measurement}

We start by reconsidering a problem that has been recently explicitly solved with the help of a slightly
different approach \cite{BFH}. We assume that the density operators are diagonal in the canonical
representation, i. e.
\begin{equation}
\label{rho-diag} \rho_1 = \sum_{i=1}^d r_{i} |r_i\rangle\langle r_i|, \qquad \rho_2 = \sum_{i=1}^d
s_{i}|s_i\rangle\langle s_i|,
\end{equation}
where  Eqs. (\ref{basis1a}) and (\ref{basis1b}) hold for the eigenstates of the density operators. The
  fidelity is then readily calculated from Eq. (\ref{F}) as
\begin{equation}
\label{fid-diag}
 F = \sum_{i=1}^d C_i \sqrt{r_is_i},
 \end{equation}
 and Eqs. (\ref{cond3}) and (\ref{cond4}) take the form
\begin{eqnarray}
\label{cond7}
   \sum_{i=1}^d \left(S_i^2 \alpha_{i} - 1 + \sqrt{\frac{\eta_2s_i}{\eta_1r_i}}C_i\right)r_i &=& 0,\\
    \label{cond8}
  \sum_{i=1}^d \left(S_i^2 \beta_{i} - 1 +  \sqrt{\frac{\eta_1r_i}{ \eta_2s_i}}C_i\right) s_i&=& 0.
   \end{eqnarray}
A solution for the diagonal elements of the optimum detection operators can now be immediately read out. It is
given by $\alpha_{i}=\alpha_i^{\rm o}$ and $\beta_{i}=\beta_{i}^{\rm o}$, where
\begin{equation}
\label{opt1} \alpha_{i}^{\rm o}=\frac{1}{S_i^2}\left(1-\sqrt{\frac{\eta_2s_i}{\eta_1r_i}}C_i\right),\;\;
\beta_{i}^{\rm o}=\frac{1}{S_i^2}\left(1-\sqrt{\frac{\eta_1r_i}{\eta_2s_i}}C_i\right).
 \end{equation}
According to Eq. (\ref{Q2}) the failure probability $Q$ does not depend on the nondiagonal elements of the
detection operators when $r_{ij}=r_i\delta_{ij}$ and $s_{ij}=s_i\delta_{ij}$. We therefore conclude that in the
optimum measurement
\begin{equation}
\label{opt}
 \alpha_{ij}=\alpha_i \delta_{ij},\quad \beta_{ij}=\beta_{i}\delta_{ij},
\end{equation}
since this requirement guarantees that $\alpha_i$ and $\beta_{i}$ can be made as large as possible while $\Pi_0$
is still a positive operator, i. e. that the failure probability becomes as small as possible.
 Because of the condition on the eigenvalues of the detection
operators we have to require that $0\leq \alpha_{i}^{\rm o}, \beta_{i}^{\rm o}\leq 1$. Therefore Eqs.
(\ref{opt1}) only represent a physical solution for the optimum measurement when the ratio $\eta_2/\eta_1$ falls
within  certain intervals. After replacing the coefficients $\alpha_i^{\rm o}$ and $\beta_i^{\rm o}$ outside
these intervals by their values at the boundaries, in order to make $Q$ as small as possible, we arrive at
\begin{equation}
\label{coeff}
\begin{array}{ll} \alpha_i^{\rm opt}=1, \quad \,\beta_i^{\rm opt}=0\;\; & \mbox{if $\;\;\quad
             \sqrt{\frac{\eta_{2}}{\eta_{1}}}\leq C_i \sqrt{\frac{r_i}{s_i}}$}, \\
\alpha_i^{\rm opt}=\alpha_i^{\rm o},\;\;\beta_i^{\rm opt}=\beta_i^{\rm o}\;\;  & \mbox{if $\;\; C_i
\sqrt{\frac{r_i}{s_i}}\leq \sqrt{\frac{\eta_{2}}{\eta_{1}}}
\leq \frac{1}{C_i}\sqrt{\frac{r_i}{s_i}}$}, \\
\alpha_i^{\rm opt}=0,\quad \;\beta_i^{\rm opt}=1\;\;& \mbox{if $\;\;
        \frac{1}{C_i}\sqrt{\frac{r_i}{s_i}}\leq \sqrt{\frac{\eta_{2}}{\eta_{1}}} $},
 \end{array}
\end{equation}
in accordance with Ref. \cite{BFH}. The optimum detection operators are then given by
\begin{equation}
\label{detect1} \Pi_1^{\rm opt} =\sum_{i=1}^{d} \alpha_{i}^{\rm opt}|v_i\rangle\langle v_i|, \quad
 \Pi_2^{\rm opt} =\sum_{i=1}^{d} \beta_{i}^{\rm opt}|w_i\rangle\langle w_i|,
\end{equation}
and
\begin{eqnarray}
\label{pi01} \Pi_0^{\rm opt} = \sum_{i=1}^d [(1 - \alpha_{i}^{\rm opt}S_i^2)|r_i\rangle\langle r_i|
                   +\alpha_{i}^{\rm opt}S_i C_i |r_i\rangle\langle w_i|\nonumber\\
                \;\;  + \alpha_{i}^{\rm opt}S_i C_i|w_i\rangle\langle r_i|
                   +(1- \alpha_{i}^{\rm opt}C_i^2-\beta_{i}^{\rm opt})|w_i\rangle\langle w_i|],\nonumber\\
\end{eqnarray}
where in the latter expression  Eqs. (\ref{pi0}) and (\ref{opt}) have been used. In order to show that these
operators indeed describe a physical measurement, we still have to verify that $\Pi_0$ is a positive operator.
  From Eq. (\ref{pi01}) it
becomes obvious that $\Pi_0$ can be represented by a matrix which consists of $d$ decoupled two by two matrices.
Taking into account that $S_i^2\alpha_i^o\beta_i^o= \alpha_i^o + \beta_i^o$, we find after minor algebra that
for each of these matrices one eigenvalue is zero and the other is given by
\begin{equation}
\label{eig2} \lambda_i= \alpha_i^o + \beta_i^o\qquad
\mbox{if $\;\; C_i \sqrt{\frac{r_i}{s_i}}\leq \sqrt{\frac{\eta_{2}}{\eta_{1}}} \leq
\frac{1}{C_i}\sqrt{\frac{r_i}{s_i}}$},
\end{equation}
or by $\lambda_i=1$ otherwise  \cite{BFH}. It is easy to check that the condition $0\leq \lambda_i\leq1$ is
indeed fulfilled for the eigenvalues $\lambda_i$ of the operator $\Pi_0$.

A few direct conclusions can be drawn from the  Eqs. (\ref{coeff}). Obviously, when for the given prior
probabilities of the two mixed states there does not exist a single value of $i$ for which the condition in the
middle line of Eq. (\ref{coeff}) is fulfilled, then the optimum measurement is a von Neumann measurement, where
the detection operators are projectors. In this case the failure probability of the optimum measurement is given
by
\begin{equation}
\label{Q-neumann}
\begin{array}{ll} Q_{\rm opt}=  1- \eta_1 \sum_{i=1}^d S_i^2 r_i \;\; & \mbox{if $\;\;
             \sqrt{\frac{\eta_{2}}{\eta_{1}}}\leq {\rm Min}_i \,\left\{C_i \sqrt{\frac{r_i}{s_i}}\right\}$}, \\
Q_{\rm opt}=  1- \eta_2 \sum_{i=1}^d S_i^2 s_i \;\; & \mbox{if $\;\;
             \sqrt{\frac{\eta_{2}}{\eta_{1}}}\geq {\rm Max}_i \left\{ \frac {1}{C_i} \sqrt{\frac{r_i}{s_i}}\right\}.$}
 \end{array}
\end{equation}
In all other cases the optimum measurement is a generalized measurement, but only when the condition in the
middle line of Eq. (\ref{coeff}) is fulfilled for each single value of $i$, $(i=1,\ldots,d)$, the fidelity bound
of the failure probability is obtained. Thus we have that $Q_{\rm opt}= 2\sqrt{\eta_1 \eta_2}F$ if
\begin{equation}
\label{cond-fid}  \mbox{ ${\rm Max}_i \,\left\{C_i \sqrt{\frac{r_i}{s_i}}\right\}\leq
\sqrt{\frac{\eta_{2}}{\eta_{1}}}\leq {\rm Min}_i \left\{ \frac {1}{C_i} \sqrt{\frac{r_i}{s_i}}\right\}.$}
\end{equation}
Clearly, when ${\rm Max}_i \,\left\{C_i \sqrt{\frac{r_i}{s_i}}\right\}\geq {\rm Min}_i \left\{ \frac {1}{C_i}
\sqrt{\frac{r_i}{s_i}}\right\}$ the condition given by  Eq. (\ref{cond-fid}) can never hold true and the overall
lower bound of the failure probability cannot be reached.

It is important to observe that the solution expressed by Eqs. (\ref{opt}) and (\ref{coeff}) holds whenever the
fidelity takes the form given by Eq. (\ref{fid-diag}), since due to Eq. (\ref{opt}) the nondiagonal density
matrix elements $r_{ij}$ and $s_{ij}$ do not enter the Eqs. (\ref{cond3}) and (\ref{cond4}). In Sec. III we
apply this solution to the optimum unambiguous discrimination of two particular density operators that do not
have to be diagonal in the canonical representation.

\section{Discrimination of states belonging to a class of similar states}

\subsection{The canonical representation and the fidelity}

Now we turn our attention to the unambiguous discrimination of two mixed states $\rho_1$ and $\rho_2$ of rank
$d$ that are connected via a unitary transformation in the $2d$-dimensional Hilbert space spanned by their joint
supports,
\begin{equation}
\label{uni}
 \rho_2 = U\;\rho_1\; U^{\dag},
\end{equation}
where $U^{\dag}=U^{-1}$. Since we want to determine the optimum measurement by means of applying Eqs.
(\ref{cond3}) and (\ref{cond4}), we first have to express the condition on the states within the framework of
the canonical representation. By inserting the respective density operators, given by Eqs. (\ref{rho1}), into
Eq. (\ref{uni}), we obtain
\begin{equation}
\label{uni1}  \rho_2=\sum_{i,j=1}^{d} s_{ij} |s_i\rangle\langle s_j| = \sum_{i,j=1}^{d}
r_{ij}U|r_i\rangle\langle r_j|U^{\dag},
\end{equation}
where
 $\langle r_i|s_j\rangle = C_i\delta_{ij}$
and  $\langle s_i|s_j\rangle = \langle r_i|r_j\rangle = \delta_{ij}$. The operator $U$ transforms any state in
the support of $\rho_1$ into a state in the support of $\rho_2$ which means in particular that $U |r_i\rangle =
\sum_k c_{ik} |s_k\rangle$, where $\sum_k c_{ik} c_{jk}^{*}= \langle r_i|r_j\rangle = \delta_{ij}$. In general,
the calculation of the fidelity of these two mixed states is a difficult problem and cannot be performed
analytically. In the following we therefore restrict ourselves to a special class of unitary transformations.

We assume that the unitary transformation $U$  can  be
 decomposed into $d$ independent unitary transformations $U_i$  that act in the $d$ mutually orthogonal
 two-dimensional subspaces spanned by the pairs of nonorthogonal states $|r_i\rangle$ and $|s_i\rangle$.
In each of the subspaces a particular orthonormal basis is given by the states $|r_i\rangle$ and
$|{w}_i\rangle$, where
\begin{equation}
\label{w}
 |{w}_i\rangle = \frac{1}{S_i} \left(|s_i \rangle - C_i  |r_i\rangle \right).
\end{equation}
Since according to Eqs. (\ref{basis1a}) and (\ref{basis1b}) the inner products of any two states in the combined
set of the basis states of the two density operators are real, the class of transformations we consider is
described by \cite{koashi}
\begin{equation}
 \label{rot}
  U = U_1(\theta_1) \otimes U_2(\theta_2) \otimes \ldots \otimes U_d(\theta_d),
\end{equation}
where the transformations in the subspaces are rotations by the angle $\theta_i$,
\begin{equation}
  U_i(\theta_i)={\rm exp}\;[\,\theta_i\,(|w_i\rangle\langle r_i|- |r_i\rangle\langle w_i|)\,].
 \end{equation}
As can be verified by expanding $U_i(\theta_i)$ in terms of powers of $\theta_i$, this is equivalent to
\begin{equation}
\label{rot2} U_i(\theta_i)|r_j\rangle =
 \left \{ \begin{array}{ll}  \cos \theta_i
|r_i\rangle +  \sin\theta_i |{w}_i\rangle\;\; & \mbox{if $ i=j$} \\
|r_j\rangle \;\; & \mbox{if $ i\neq j$}.
 \end{array}
\right.
\end{equation}

In order to obtain the canonical representation of the density operators, we have to determine the eigenvalues
and eigenstates of the operator $P_1 P_2 P_1$,  see Eq. (\ref{basis1}).
 The projectors onto the supports of $\rho_1$ and $\rho_2$ read
\begin{equation}
\label{P2}  P_1 = \sum_{i=1}^{d} |r_i\rangle\langle r_i|,\quad   P_2  = \sum_{i=1}^{d} U|r_i\rangle\langle
r_i|U^{\dag},
\end{equation}
where the expression for $P_2$ follows from the right-hand side of  Eq. (\ref{uni1}). By applying Eq.
(\ref{rot2}) we easily find that
\begin{equation}
\label{P1P2}  P_1 P_2 P_1  = \sum_{i=1}^{d} \cos^2 \theta_i |r_i\rangle\langle r_i|,
\end{equation}
and  Eq. (\ref{basis1}) therefore immediately yields
\begin{equation}
\label{abbr1}  C_i = \cos\theta_i.
\end{equation}
From  Eqs. (\ref{rot2}) and  Eq. (\ref{w}) we then obtain
\begin{equation}
U |r_i\rangle =|s_i\rangle
\end{equation}
which means that
\begin{eqnarray}
\label{uni3}
 \langle r_j| U |r_i\rangle &=&  C_i \delta_{ij}.
 \end{eqnarray}
 After calculating the matrix element $\langle r_i |\rho_2|r_j\rangle$ from both expressions in Eq.
(\ref{uni1}), using Eq. (\ref{uni3}), we finally get
\begin{equation}
\label{diag}
 s_{ij}= r_{ij}.
\end{equation}
Hence under the condition given by Eq. (\ref{rot}) our starting equation,  Eq . (\ref{uni1}), can only be
fulfilled when
\begin{equation}
\label{uni2}  \rho_1= \sum_{i,j=1}^{d} r_{ij} |r_i\rangle\langle r_j|, \quad \rho_2= \sum_{i,j=1}^{d} r_{ij}
 |s_i\rangle\langle s_j|.
\end{equation}
In other words, the two mixed states we consider differ by the orientation of their respective canonical basis
states in the $2d$-dimensional Hilbert space, but the relative weights of these states and the coherences
between them are the same.

After having specified the relation between the matrix elements of the two density operators, our next step
before applying Eqs. (\ref{cond3}) and (\ref{cond4}) is the calculation of the fidelity. From Eq. (\ref{uni}) we
obtain $\sqrt{\rho_2}= U \sqrt{\rho_1}U^{\dag}$ and Eq. (\ref{F}) therefore yields
\begin{eqnarray}
\label{fid} F&=& {\rm Tr}|\sqrt{\rho_1}U \sqrt{\rho_1}U^{\dag}|\nonumber\\
 &=& {\rm Tr}[(\sqrt{\rho_1}U
\sqrt{\rho_1}U^{\dag} \;
 U \sqrt{\rho_1}U^{\dag} \sqrt{\rho_1})^{\frac{1}{2}}]\nonumber\\
 &=& {\rm Tr} |\sqrt{\rho_1} U \sqrt{\rho_1}|,
\end{eqnarray}
where we made use of the fact that $U^{\dag}U=I$. Writing the unity operator in our 2$d$-dimensional Hilbert
space as $ I=\sum_{i=1}^d(|r_i\rangle \langle r_i|+ |w_i\rangle \langle w_i|)$ and inserting it twice, taking
into account that $\rho_1|w_i\rangle =0$, we obtain
\begin{eqnarray}
\label{fid1} F
 &=& {\rm Tr}|\sqrt{\rho_1}\sum_{i,j}|r_i\rangle \langle r_i|U |r_j\rangle
\langle r_j|\sqrt{\rho_1}|\nonumber\\
 &=& {\rm Tr}|\sum_{i}C_i \sqrt{\rho_1} |r_i\rangle \langle r_i|\sqrt{\rho_1}|,
 \end{eqnarray}
where Eq. (\ref{uni3}) has been used. Defining the vector $|a_i\rangle =  \sqrt{\rho_1} |r_i\rangle$,
we find that $F=\sum_i C_i {\rm Tr}(|a_i\rangle \langle a_i|)$ and arrive at the final result
\begin{equation}
\label{fid2}  F= \sum_{i=1}^d C_i  \langle r_i|\rho_1 |r_i\rangle = \sum_{i=1}^d C_i r_i.
\end{equation}
Interestingly, for the class of states we consider the fidelity does not depend on the nondiagonal elements of
the density operators in the canonical representation, no matter what is the kind of the individual unitary
transformations in the two-dimensional subspaces.

\subsection{The optimum measurement}

We are now prepared  to determine the measurement for the optimum unambiguous discrimination.
Upon inserting the
expression for the fidelity, Eq. (\ref{fid2}), into our basic conditions,
 Eqs. (\ref{cond3}) and (\ref{cond4}), taking into account that  $s_{ij}=r_{ij}$, we arrive at
the two equations
\begin{eqnarray}
\label{cond9}
 \sum_{i\neq j}S_iS_j \alpha_{ij}r_{ji}+ \sum_i \left(S_i^2 \alpha_{i}
 - 1 + \sqrt{\frac{\eta_2}{\eta_1}}C_i\right)r_i &=& 0,\nonumber\\
    \label{cond10}
 \sum_{i\neq j}S_iS_j \beta_{ij}r_{ji}+ \sum_i \left(S_i^2 \beta_{i}
 - 1 +  \sqrt{\frac{\eta_1}{ \eta_2}}C_i\right) r_i&=& 0\nonumber\\
   \end{eqnarray}
that have to be fulfilled by the coefficients determining the optimum detection operators. Because of the
special structure of these equations, resulting from the specific expression for the fidelity, we are free to
make the Ansatz
\begin{equation}
\label{opt2}
 \alpha_{ij}=\alpha_i \delta_{ij},\quad \beta_{ij}=\beta_{i}\delta_{ij}.
\end{equation}
Obviously, the problem to be solved is then reduced to the problem expressed by Eqs. (\ref{cond7}) and
(\ref{cond8}), in the special case that $r_i=s_i$. The previous solution, given by Eqs. (\ref{opt1}) -
(\ref{coeff}),  therefore can be immediately applied and the optimum coefficients read
\begin{equation}
\label{coeff2}
\begin{array}{ll} \alpha_i^{\rm opt}=1, \quad \,\beta_i^{\rm opt}=0\;\;\quad & \mbox{if $\;\;\quad
             \sqrt{\frac{\eta_{2}}{\eta_{1}}}\leq C_i, $} \\
\alpha_i^{\rm opt}=\alpha_i^{\rm o},\;\;\beta_i^{\rm opt}=\beta_i^{\rm o}\;\; \quad & \mbox{if $\;\; C_i \leq
\sqrt{\frac{\eta_{2}}{\eta_{1}}}
\leq \frac{1}{C_i}$}, \\
\alpha_i^{\rm opt}=0,\quad \;\beta_i^{\rm opt}=1\;\;& \mbox{if $\;\;
        \frac{1}{C_i}\leq \sqrt{\frac{\eta_{2}}{\eta_{1}}} $},
 \end{array}
\end{equation}
where
\begin{equation}
\label{opt3} \alpha_{i}^{\rm o}=\frac{1}{S_i^2}\left(1-\sqrt{\frac{\eta_2}{\eta_1}}C_i\right),\;\;
\beta_{i}^{\rm o}=\frac{1}{S_i^2}\left(1-\sqrt{\frac{\eta_1}{\eta_2}}C_i\right).
 \end{equation}
The solutions  for the optimum detection operators follow by inserting the optimum coefficients into Eqs.
(\ref{detect1}) and (\ref{pi01}).

In order to obtain compact results for the minimum failure probability, $Q_{\rm opt}$,
ensuing from the optimum
measurement, it will be useful to adopt the convention that
\begin{equation}
\label{order} C_1\leq C_2 \leq\ldots \leq C_{d-1}\leq C_d.
 \end{equation}
After inserting  Eqs. (\ref{opt2}) - (\ref{opt3}) into the equation for the failure probability $Q$,  Eq.
(\ref{Q2}), taking into account that $r_i=s_i$, we find again that the structure of the resulting expressions
depends on the ratio of the prior probabilities. If the latter is such that one of the two von Neumann
measurements is optimal, the minimum failure probability takes the form
\begin{equation}
\label{Q-uniform1}
\begin{array} {ll}Q_{\rm opt}=  1- \eta_1 \sum_{i=1}^d S_i^2 r_i \;\;\quad & \mbox{if $\;\;
             \sqrt{\frac{\eta_{2}}{\eta_{1}}}\leq C_1$}, \\
Q_{\rm opt}=  1- \eta_2 \sum_{i=1}^d S_i^2 r_i \;\;\quad & \mbox{if $\;\;
             \sqrt{\frac{\eta_{2}}{\eta_{1}}}\geq \frac {1}{C_1} $}.
 \end{array}
\end{equation}
On the other hand, with respect to the saturation of the fidelity bound we find that
\begin{equation}
\label{Q-uniform2} Q_{\rm opt}=  2\sqrt{\eta_1\eta_2}F\;\;\quad\quad \mbox{if $\;\;C_d \leq
             \sqrt{\frac{\eta_{2}}{\eta_{1}}}\leq \frac {1}{C_d} $},
 \end{equation}
where $F=\sum_{i=1}^d C_i r_i$. In the intermediate regions of the ratio of the prior probabilities the optimum
failure probability can be written as
\begin{eqnarray}
\label{Q-uniform3} Q_{\rm opt}&=&1-\sum_{i=1}^k (1-2\sqrt{\eta_1\eta_2}C_i)r_i - \eta_1 \sum_{i=k+1}^d S_i^2
r_i\\
 &&\mbox{ if $\;\; C_k \leq \sqrt{\frac{\eta_2}{\eta_1}} \leq C_{k+1}\quad(1\leq k\leq
 d-1)$}\,\quad\qquad\nonumber
 \end{eqnarray}
and
\begin{eqnarray}
\label{Q-uniform4} Q_{\rm opt}&=&1-\sum_{i=1}^k (1-2\sqrt{\eta_1\eta_2}C_i)r_i - \eta_2 \sum_{i=k+1}^d S_i^2
r_i\\
&& \mbox{ if $\;\; \frac{1}{C_{k+1}} \leq \sqrt{\frac{\eta_2}{\eta_1}}
 \leq \frac{1}{C_k}\quad(1\leq k\leq d-1)$}.\;\quad\qquad\nonumber
 \end{eqnarray}
Clearly, in dependence on the ratio of the prior probabilities of the states, there are in general 2$d$+1
parameter regions in which the optimum measurement operators have a different structure and consequently the
expression for the optimum failure probability takes a different form. These regions do not depend on the matrix
elements of the density operators, but only on the canonical angles, that is on the constants $C_i$.  For $d=2$,
the calculation of $C_1$ and $C_2$ can be easily performed analytically by means of Eq. (\ref{basis1}) since it
only amounts to the solution of a quadratic equation.

It is interesting to compare the parameter interval in which the fidelity bound of the failure probability can
be actually achieved, specified in Eq. (\ref{Q-uniform2}), with the respective parameter interval following from
a necessary, but not sufficient condition \cite{HB}, as given in Eq. (\ref{ness-cond}). Representing $P_1$ and
$P_2$ as $\sum_{i=1}^d |r_i\rangle\langle r_i|$ and $\sum_{i=1}^d |s_i\rangle\langle s_i|$, respectively, we
find that ${\rm Tr}(P_2\rho_1)={\rm Tr}(P_2\rho_1)=\sum_{i=1}^d C_i^2 r_i$. The former interval is necessarily
not larger than the latter, the relative difference between the intervals obviously being
characterized by the ratio
$\sum_{i=1}^d C_i^2 r_i/\sum_{i=1}^d C_i C_d r_i$, where the explicit expression for the fidelity has been taken
into account.

Two special cases are worth mentioning. In the first one the two mixed states have equal prior probabilities to
occur, $\eta_1=\eta_2=0.5.$ Since the inequality $C_d\leq 1 \leq 1/C_d$ certainly holds for any $C_d =
\cos\theta_d\leq 1 $, it becomes obvious from Eq. (\ref{Q-uniform2}) that in this case the fidelity bound of the
failure probability can always be reached.

The second special case refers to identical canonical angles,
$C_i=\cos\theta$ for $i=1,\ldots,d$
which means that the two density operators are connected via a rotation by the angle $\theta$. We mention that
for a nonorthogonal angle $\theta$ this is exactly the condition that has been derived in Ref. \cite{koashi} as
the prerequisite for secure quantum communication when the two-pure-state protocol \cite{bennett} is extended to
two mixed states. In this case it follows that ${\rm Tr}(P_1\rho_2)={\rm Tr}(P_2\rho_1)=F^2 =\cos^2 \theta $ and
our general solution, represented by Eqs. (\ref{Q-uniform1}) - (\ref{Q-uniform4}) reduces to
\begin{equation}
Q_{\rm opt}= \left \{
\begin{array}{ll}
2 \sqrt{\eta_1 \eta_2}F\;\; & \mbox{if
$F\leq \sqrt{\frac{\eta_1}{\eta_2}} \leq \frac{1}{F}$}\\
 \eta_{\rm min} +
\eta_{\rm max} F^2\;\; & \mbox{otherwise},
\end{array}
\right. \nonumber
\end{equation}
where $\eta_{\rm min} (\eta_{\rm max})$ denotes the smaller (larger) of the prior probabilities. This result
exactly corresponds to the solution for the optimum unambiguous discrimination of two pure states \cite{jaeger}.

\section{Conclusions}

In this paper we have shown that an analytical solution for the optimum unambiguous discrimination of two mixed
states can be obtained provided that the expression for their fidelity is given by Eq. (\ref{fid-diag}), where
the density operators are represented with the help of the canonical basis that separates the joint Hilbert
space into $d$ mutually orthogonal two-dimensional subspaces.  The discrimination problem is then mathematically
equivalent  to distinguishing pairs of pure states. We applied the solution to the discrimination of two mixed
states that belong to a special class of similar states. The density operators of these states do not have to be
diagonal in the canonical representation. Our results might be also of interest for quantum cryptography, where
states of the kind considered in this paper play a role \cite{koashi}.

We still note that after finishing this work a related paper appeared \cite{zhou} where the authors investigate
lower bounds of the failure probability by introducing a different kind of state vectors spanning the joint
Hilbert space. These states are defined by the requirement that for any two density operators the expression for
the fidelity is of a form equivalent to Eq. (\ref{fid-diag}). In contrast to the canonical basis states
considered in the present paper, the states introduced in \cite{zhou} do not necessarily provide an orthogonal
basis in the supports of the two density operators.

\begin{acknowledgments}
The author would like to thank Janos Bergou (Hunter College, New York) and Philippe Raynal (Universit\"at
Erlangen) for useful discussions.

\end{acknowledgments}

\end{document}